\documentclass[apj]{emulateapj}
\slugcomment{{\sc Accepted to Apj:} April 7, 2010}
\usepackage{natbib}
\usepackage{lscape}
\usepackage{longtable}
\usepackage{setspace}
\def\CIVdblt{{\rm C}\kern 0.1em{\sc iv}~$\lambda\lambda 1548, 1550$}
\def\MgIIdblt{{\rm Mg}\kern 0.1em{\sc ii}~$\lambda\lambda 2796, 2803$}
\def\NVdblt{{\rm N}\kern 0.1em{\sc v}~$\lambda\lambda 1238, 1242$}  
\def\OVIdblt{{\rm O}\kern 0.1em{\sc vi}~$\lambda\lambda 1031, 1037$}
\def\SiIVdblt{{\rm Si}\kern 0.1em{\sc iv}~$\lambda\lambda1394, 1403$}
\def\AlIIIdblt{{\rm Al}\kern 0.1em{\sc iii}~$\lambda\lambda1855,1863$}
\def\FeIIdblt{{\rm Fe}\kern 0.1em{\sc ii}~$\lambda\lambda 2383, 2600$}

\def\AlII{\hbox{{\rm Al}\kern 0.1em{\sc ii}}}
\def\AlIII{{\hbox{\rm Al}\kern 0.1em{\sc iii}}}
\def\CaI{\hbox{{\rm Ca}\kern 0.1em{\sc i}}}
\def\CaII{\hbox{{\rm Ca}\kern 0.1em{\sc ii}}}
\def\CrII{\hbox{{\rm Cr}\kern 0.1em{\sc ii}}}
\def\CII{\hbox{{\rm C}\kern 0.1em{\sc ii}}}
\def\CIII{\hbox{{\rm C}\kern 0.1em{\sc iii}}}
\def\CIV{\hbox{{\rm C}\kern 0.1em{\sc iv}}}
\def\CV{\hbox{{\rm C}\kern 0.1em{\sc v}}}

\def\HI{\hbox{{\rm H}\kern 0.1em{\sc i}}}
\def\HII{\hbox{{\rm H}\kern 0.1em{\sc ii}}}
\def\Lya{\hbox{{\rm Ly}\kern 0.1em$\alpha$}}
\def\Lyb{\hbox{{\rm Ly}\kern 0.1em$\beta$}}
\def\Lyg{\hbox{{\rm Ly}\kern 0.1em$\gamma$}}
\def\Lyfive{\hbox{{\rm Ly}\kern 0.1em$5$}}
\def\Lysix{\hbox{{\rm Ly}\kern 0.1em$6$}}
\def\Lyseven{\hbox{{\rm Ly}\kern 0.1em$7$}}
\def\Lyeight{\hbox{{\rm Ly}\kern 0.1em$8$}}
\def\Lynine{\hbox{{\rm Ly}\kern 0.1em$9$}}
\def\Lyten{\hbox{{\rm Ly}\kern 0.1em$10$}}
\def\HeI{\hbox{{\rm He}\kern 0.1em{\sc i}}}
\def\HeII{\hbox{{\rm He}\kern 0.1em{\sc ii}}}
\def\FeI{\hbox{{\rm Fe}\kern 0.1em{\sc i}}}
\def\FeII{\hbox{{\rm Fe}\kern 0.1em{\sc ii}}}
\def\FeIII{\hbox{{\rm Fe}\kern 0.1em{\sc iii}}}
\def\MnII{\hbox{{\rm Mn}\kern 0.1em{\sc ii}}}
\def\MgI{\hbox{{\rm Mg}\kern 0.1em{\sc i}}}
\def\MgII{\hbox{{\rm Mg}\kern 0.1em{\sc ii}}}
\def\MgIII{\hbox{{\rm Mg}\kern 0.1em{\sc iii}}}
\def\MgIV{\hbox{{\rm Mg}\kern 0.1em{\sc iv}}}
\def\NaI{\hbox{{\rm Na}\kern 0.1em{\sc i}}}
\def\NV{\hbox{{\rm N}\kern 0.1em{\sc v}}}
\def\NII{\hbox{{\rm N}\kern 0.1em{\sc ii}}}
\def\NIII{\hbox{{\rm N}\kern 0.1em{\sc iii}}}
\def\OVI{\hbox{{\rm O}\kern 0.1em{\sc vi}}}
\def\OII{\hbox{[{\rm O}\kern 0.1em{\sc ii}]}}
\def\SiI{\hbox{{\rm Si}\kern 0.1em{\sc i}}}
\def\SiIII{\hbox{{\rm Si}\kern 0.1em{\sc iii}}}
\def\SiIV{\hbox{{\rm Si}\kern 0.1em{\sc iv}}}
\def\SII{\hbox{{\rm S}\kern 0.1em{\sc ii}}}
\def\SIII{\hbox{{\rm S}\kern 0.1em{\sc iii}}}
\def\SIV{\hbox{{\rm S}\kern 0.1em{\sc iv}}}
\def\TiII{\hbox{{\rm Ti}\kern 0.1em{\sc ii}}}
\def\ZnII{\hbox{{\rm Zn}\kern 0.1em{\sc ii}}}
\newcommand{\kms}{\hbox{km~s$^{-1}$}}
\newcommand{\cmsq}{\hbox{cm$^{-2}$}}
\newcommand{\cc}{\hbox{cm$^{-3}$}}
\def\kms{\hbox{km~s$^{-1}$}}      
\def\cmsq{\hbox{cm$^{-2}$}}
\def\cc{\hbox{cm$^{-3}$}}

\begin{document}
\title{A Bare Molecular Cloud at $z \sim 0.45$\altaffilmark{1}}
\author{Therese~M.~Jones\altaffilmark{2}, Toru~Misawa\altaffilmark{3}, Jane~C.~Charlton\altaffilmark{4}, Andrew~C.~Mshar\altaffilmark{4}, Gary~J.~Ferland\altaffilmark{5}}
\altaffiltext{1}{Based on public data obtained from the ESO archive of
observations from the UVES spectrograph at the VLT, Paranal, Chile}

\altaffiltext{2}{Department of Astronomy, University of California, Berkeley, CA 94720, USA, {\it tjones@astro.berkeley.edu}}

\altaffiltext{3}{School of General Education, Shinshu University,
  3-1-1 Asahi, Matsumoto, Nagano, 390-8621 Japan {\it misawatr@shinshu-u.ac.jp}}

\altaffiltext{4}{Department of Astronomy and Astrophysics, The Pennsylvania State University, University Park, PA 16802, USA, {\it charlton@astro.psu.edu; acmshar@gmail.com}}

\altaffiltext{5}{Department of Physics and Astronomy, University of Kentucky,
Lexington, KY 40506, USA, {\it gary@pa.uky.edu}}

\begin{abstract} 

Several neutral species ({\MgI}, {\SiI}, {\CaI}, {\FeI}) have been
detected in a weak {\MgII} absorption line system ($W_r(2796) \sim 0.15~${\AA}) at
$z \sim 0.45$ along the sightline toward HE0001-2340.  These
observations require extreme
physical conditions, as noted in \citet{D'Odorico07}.  We place
further constraints on the properties of this system by running a wide
grid of photoionization models, determining that the absorbing cloud that produces the neutral absorption is 
extremely dense ($\sim 100-1000$ {\cc}), cold ($<100$K), and has significant 
molecular content ($\sim 72-94 \%$).  Structures of this size and temperature have
been detected in Milky Way CO surveys, and have been predicted in hydrodynamic simulations
of turbulent gas.  In order to explain the observed line
profiles in all neutral and singly ionized chemical transitions, the lines
must suffer from unresolved saturation and/or the absorber must partially
cover the broad emission line region of the background quasar.  In addition
to this highly unusual cloud, three other ordinary weak {\MgII} clouds 
(within densities of $\sim 0.005$~{\cc} and temperatures of $\sim10000$~K)
lie within 500~{\kms} along the same sightline.  We suggest that
the ``bare molecular cloud'', which appears to reside outside of a galaxy
disk, may have had in situ star formation and may evolve into an ordinary
weak {\MgII} absorbing cloud. 
\end{abstract}

\section{Introduction}
\label{sec:1}

Observations of {\MgIIdblt} in intervening quasar absorption line
systems are vital to the study of the interstellar medium of low-redshift 
galaxies and their environments,
as {\MgIIdblt} lies in the optical from $z \sim 0.3-2.4$, and
serves as a probe of low-ionization gas.  Through photoionization
models, it is possible to derive many properties of the absorbing
gas.  This includes the line-of-sight extent, density, temperature, and
molecular content, which are
constrained by absorption in the different ionization states of
various chemical elements.

\citet{CWC99b} and \citet{Rigby} found that roughly one-third of weak
{\MgII} absorbers at redshifts $0.4 < z < 1.4$, with $W_r(2796) <
0.3${\AA}, are in multiple-cloud systems.  They propose that many of
the weak absorbers contain multi-phase structures with a range of
levels of ionization.  Such absorbers are generally thought to arise
in sub-Lyman limit system environments, with $N({\HI}) <
10^{17.2}~{\cmsq}$, and metallicities log$[Z/Z_{\odot}]> -1$
\citep{Churchill00}.

Strong {\MgII} absorbers ($W_r(2796) > 0.3$~{\AA}), in contrast, are
almost always associated with Lyman limit systems ($N({\HI}) > 10^{17.2}~{\cmsq}$),
and many with damped {\Lya} absorbers (DLAs; with $N({\HI}) > 2$x$10^{20}~{\cmsq}$).
\citet{Rao06} found that $36\%$ of MgII absorbers with $W_r(2796) >
0.5$~{\AA} and {\FeII} $W_r(2600) > 0.5$~{\AA} were DLAs in an HST
survey for $z<1.65$ systems.  In that sample, the
average $N({\HI})$ was $9.7 \pm 2.2$ x $10^{18}~{\cmsq}$ for $0.3 <
W_r(2796) < 0.6$~{\AA}, and $3.5 \pm 0.7$ x $10^{20}~{\cmsq} $ for $W_r(2796)
> 0.6$~{\AA}.  Most DLAs at low $z$ are thought to be associated with galaxies
with a variety of morphological types, from $0.1 L^*$ galaxies to low
surface brightness galaxies \citep{LeBrun97, RT98,
Bowen01, RT03}. \citet{Ledoux03} find molecular hydrogen in 13-20\% of DLA
systems at high redshift, but note that there is no correlation
between the detection of molecular hydrogen and {\HI} column density.
Despite this lack of correlation, \cite{Petitjean06} find molecular hydrogen in 9 out of 18 high metallicity systems ([X/H]$>-1.3$) at high redshift.

In this paper, we study a multiple-cloud weak {\MgII} system toward
HE0001-2340 ($z_{em}=2.28$, \citealt{Reimers98}) at $z= 0.4524$.  The
four weak {\MgII} clouds are spread over a velocity range of $\sim
600$~{\kms}. {\MnII} and {\CaII}, which are generally only detected in
the very strongest absorbers, are found in one of the four clouds
comprising this system. {\FeI}, {\SiI}, and {\CaI} are also detected
in this cloud; these neutral states have not been reported to exist in
any other extragalactic environment, even in most DLAs, and have only
been found in several dense galactic molecular clouds \citep{Welty03}.
\citet{D'Odorico07} notes that the ratios of ${\MgI}/{\MgII}$ and
${\CaI}/{\CaII}$ in this system are orders of magnitude higher than in
other absorbers, implying a very low ionization state.  She also
observes that there is an extreme underabundance of Mg with respect to
Fe, which cannot be explained by nucleosynthesis or dust depletion,
and cannot be reproduced by photoionization models.  The metallicity
of the system cannot be directly determined from {\Lya} due to the
absorption from a Lyman limit system at $z=2.18$ \citep{Reimers98}.

Due to a lack of metallicity constraints, \citet{D'Odorico07} assumes
DLA-regime column densities, noting that the observed $N({\MgI})$ of
the system is comparable to the sample of 11 DLA by \citet{DZ06}.  She
further constrains her parameters by comparing to the local cold
interstellar clouds of \citet{Welty03}, finding that systems with
similar amounts of ${\FeI}$ have metallicities $-3.78 <
[\rm{Fe/H}]]<-2.78$.

With such a low assumed metallicity and high ${\HI}$ column density,
\citet{D'Odorico07} is unable to reproduce the observed ratios of
$\frac{N({\MgI})}{N({\MgII})}, \frac{N({\CaI})}{N({\CaII})}$ and
$\frac{N({\FeI})}{N({\FeII})}$ in photoionization models, 
and is thus unable to draw concrete
conclusions about the properties of this cloud.   
We propose that these noted differences
suggest that this unusual cloud is part of class of systems unrelated
to previously observed DLAs, which \citet{D'Odorico07} notes have
$N({\MgII})$ two orders of magnitude
higher than this system, as well as no previous {\FeI}, {\CaI}, and
{\SiI} detections.  We therefore explore a range of metallicities and
$N({HI})$ in our modeling process.

As it is difficult to reproduce the line ratios in this unusual cloud,
we also expand the consideration of parameter space to
explore the possibility of unresolved saturation.  The effect
of partial covering of the background quasar is also explored, since
photoionization models of the system suggest the cloud is compact
enough to partially cover the broad emission line region of the
quasar, with high densities ($n_H=1-34$ {\cc}), cold
temperatures ($<200$K), and a molecular hydrogen fraction larger than
$20\%$.  Partial covering has only been observed once before in an
intervening quasar absorption line system. In the lensed quasar APM
08279+5255 at $z=3.911$ \citep{Kobayashi02}, one-third of the
components of a strong {\MgII} absorber were not detected toward the
second image of the QSO, while fits to the components suggested
$C_f=0.5$ in the other image, constraining the absorber size to be as
small as $200$~pc. 

We begin, in \S\ref{sec:2}, with a description of the VLT/UVES
spectrum of HE~$0001-2340$, and display and quantify the observed
properties of the $z=0.4524$ multiple cloud weak {\MgII} system.
\S\ref{sec:3} details the Voigt profile fit performed on this
four-cloud system, including covering factor analysis of the first
cloud.  It also describes the photoionization modeling method used to
constrain the ionization parameters/densities of the four
clouds. \S\ref{sec:4} gives modeling results for each cloud, while
\S\ref{sec:5} discusses the implications of the cloud models on the 
origin of the gas, while \S\ref{sec:6} summarizes the findings and 
considers the properties of the absorption system in the context of
broader questions relating to galaxy environments.

\section{The $z=0.4524$ Absorber Toward HE~0001-2340}
\label{sec:2}
A spectrum of HE0001-2340, taken in 2001, was procured from the ESO
archive, having been obtained as
part of the ESO-VLT Large Programme ``The Cosmic Evolution of the IGM'' 
\citep{Richter05}.
This $z_{em}=2.28$ quasar, HE0001-2340 ($V=16.7$), was observed with
the Ultraviolet and Visual Echelle Spectrograph (UVES) on the
Very Large Telescope (VLT), as detailed in \citet{Richter05}.
The data were reduced as described in \citet{Kim04}.  The spectrum has a 
resolution R $\sim 45,000$ ($\sim 6.6$ {\kms}) 
and covers a range of 3050-10070~{\AA}.  Breaks in wavelength coverage
occurred at 5750-5840~{\AA} and 8500-8660~{\AA}.
The spectrum is of extremely high quality, with $S/N > 100$ per pixel
over most of the wavelength coverage. 
Continuum fitting (with a cubic spline function) was performed using 
the IRAF\footnotemark[5] SFIT task as described in \citet{Lynch06}.

\footnotetext[5]{IRAF is distributed by the National Optical Astronomy
 Observatories which are operated by AURA, Inc., under contract to the National
 Science Foundation.}

\subsection{The System}
\label{sec:2.1}

\begin{figure}
\centering
\epsscale{1.0}
\plotone{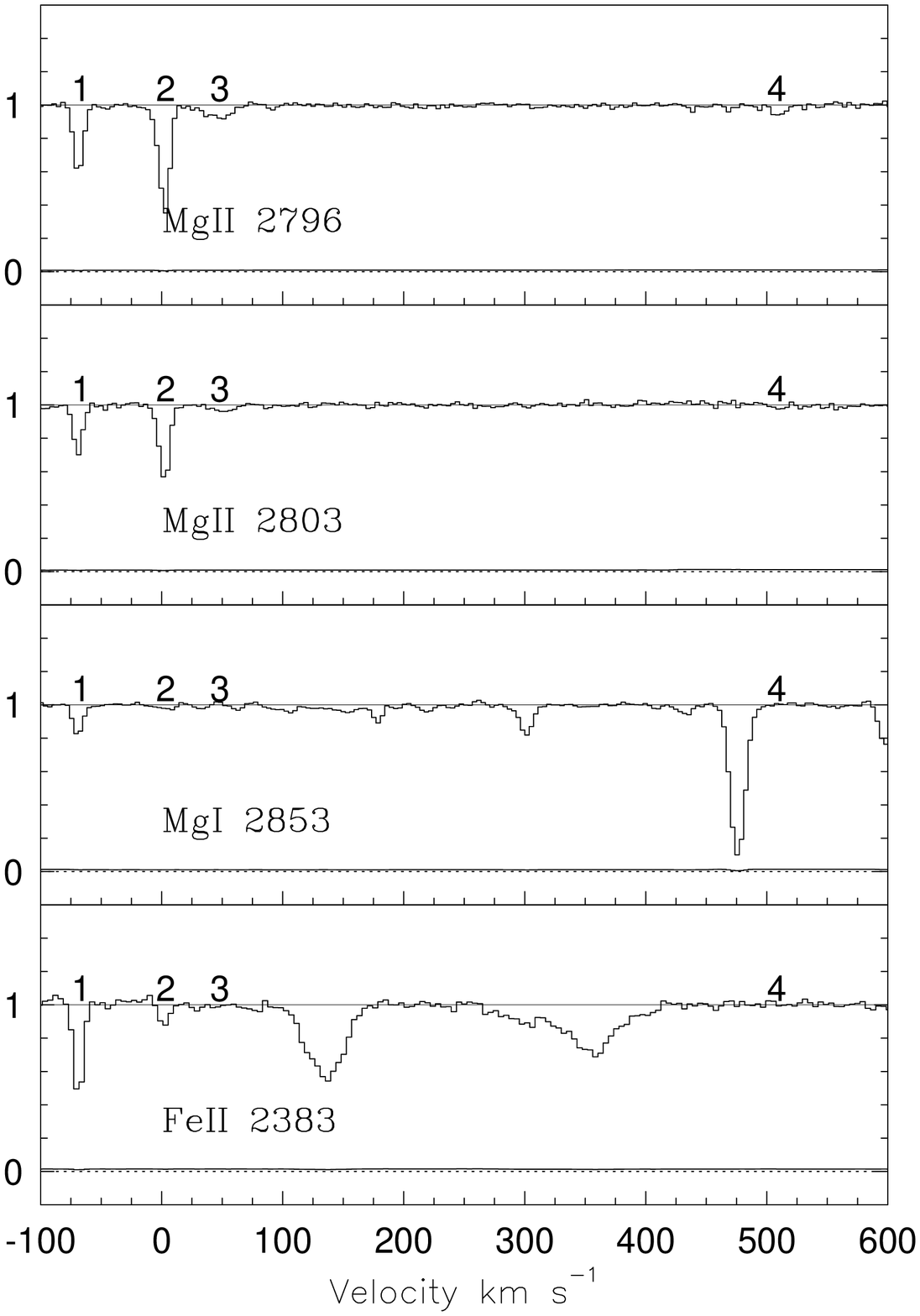}
\caption
{Four detected clouds at z $\sim$ 0.4524 along the line of sight toward HE0001-2340.
The zero point in velocity is set at the optical depth weighted mean velocity
of the whole system, at $z=0.452399$.  {\FeII}~$\lambda$~2383 is detected in Clouds 1 and 2, and the {\FeII} to {\MgII} ratio in Cloud 1 is remarkable.  
\label{fig:clouds_labeled}}
\end{figure}

\begin{figure}
\centering
\epsscale{1.0}
\plotone{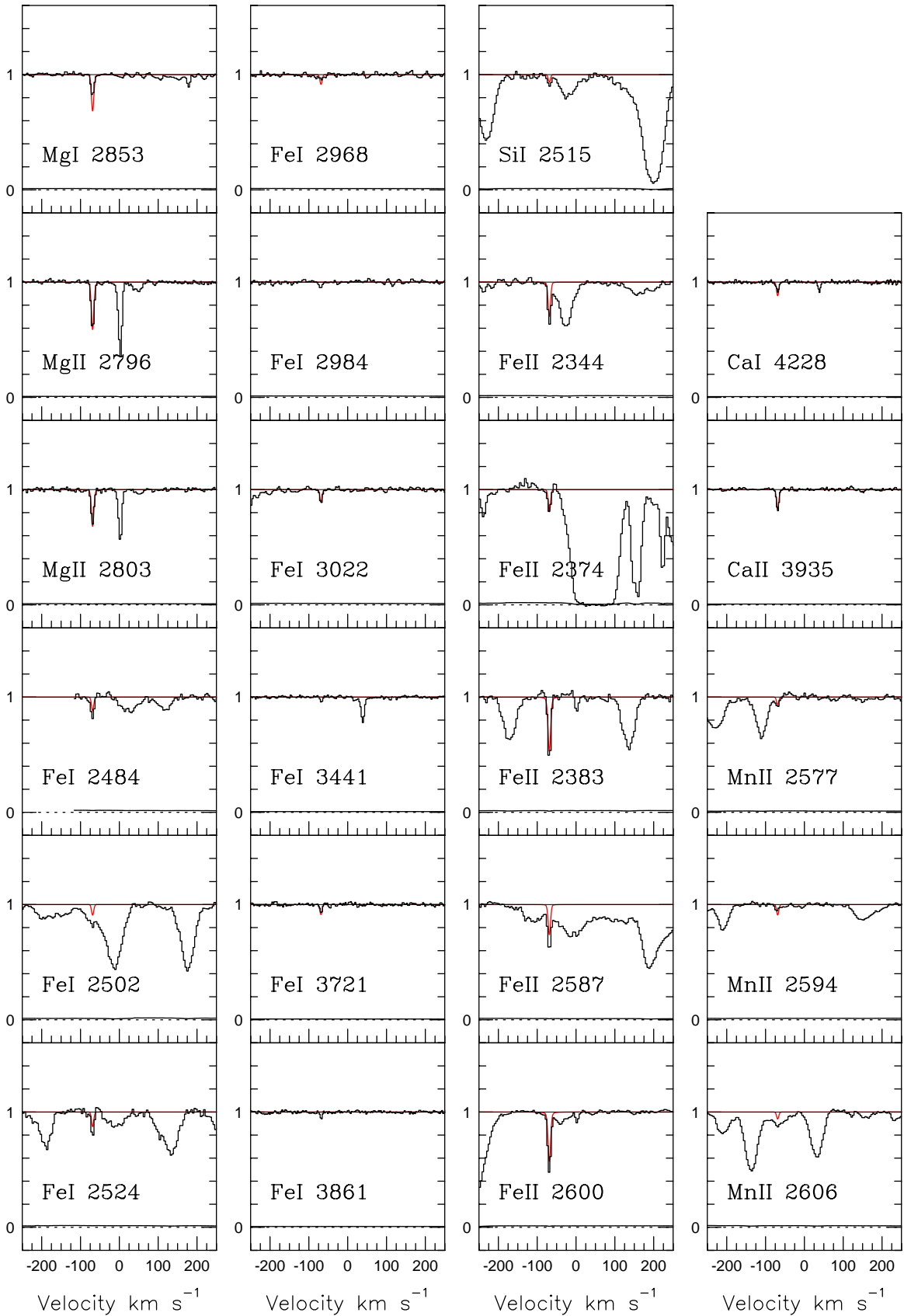}
\caption
{VLT spectrum (black) and photoionization model (red) in which Cloud 1 is modeled
including unresolved saturation, and with $\log Z=-1.0$, $\log U=-8.2$,
$\log(N${\MgII}$))=14.1$, and $b=0.2${\kms}. 
\label{fig:system_plot_unresolved}}
\end{figure}

In the $z=0.4524$ system, we detect four distinct {\MgII}~$\lambda$2796 subsystems
at $>5
\sigma$ levels in the VLT spectrum, and all are confirmed by
{\MgII}~$\lambda$2803.  Fig.~\ref{fig:clouds_labeled} shows the
location of these features, with a velocity scale centered so that half
of the system optical depth lies blueward of $0$~{\kms}, at
$z=0.452399$.  Detections may be noted at $-69$, $0$, $47$, and $507$ {\kms}
in {\MgIIdblt}, at $-69$ and $0$ {\kms} in {\FeII}, and at $-69$ {\kms} in
{\MgI}.  The fourth cloud is very weak, with {\MgII}~$\lambda$2803
detected at just over $3 \sigma$; this detection is made possible by
the very high S/N of the spectrum.  The first cloud also has detected {\FeI}, {\SiI}, {\CaI}, {\CaII}, and {\MnII}, as seen in
Figure ~\ref{fig:system_plot_unresolved}.

All transitions blueward of $\sim4000$~{\AA} ($\sim2750$~{\AA} in the
rest frame of the $z=0.4524$ system) are potentially contaminated by {\Lya} forest
lines.  The only two detected features displayed in
Fig.~\ref{fig:system_plot_unresolved} that are not from the $z=0.4524$ system, and are
outside of the forest region, occur in the {\CaI}~$\lambda$4228 panel,
at $\sim30$~{\kms}, and in the {\FeI}~$\lambda$3022 panel, at
$\sim270$~{\kms}.  The former feature is {\FeII}~$\lambda$2374 from a
system at $z=1.587$, while the latter is {\CIV}~$\lambda$1548 from a
system at $z=1.838$.  Lyman series lines for the $z=0.4524$ system are
unavailable, due to a full Lyman Limit break from a system at
$z=2.187$ \citep{Richter05}, allowing no direct constraints on the
metallicity of the system.

Rest frame equivalent widths for all transitions detected at $>3 \sigma$
in the spectra, as determined by Gaussian fits to the unresolved line
profiles, are given in Table~\ref{tab:ew} for the strongest transitions.
The column densities and Doppler parameters, using Voigt profile fitting,
assuming full coverage,
are listed in Table~\ref{tab:Nb}.  Cloud 2 was resolved into two blended
components by our fitting procedure.
Table~\ref{tab:ew2} shows equivalent widths for additional transitions
detected in Cloud 1.  Upper limits are given for blended
detections.

Since the detection of neutral transitions in a weak {\MgII} absorber
is quite surprising, we must consider whether the apparent {\FeI},
{\CaI}, {\SiI}, and {\MgI} detections for Cloud 1 are valid, and not
chance superpositions.  In Table~\ref{tab:ew2}, the oscillator
strengths of the nine detected {\FeI}
transitions are listed, along with their equivalent widths.
Only three of these detections, {\FeI}$\lambda$2484,
$\lambda$2502, and $\lambda$2524, are in the {\Lya} forest. 
The oscillator strengths of
the various transitions are roughly consistent with the relative
equivalent widths, however, a detailed analysis suggests that partial
covering or unresolved saturation affects the line profiles, as
discussed further in \S~\ref{sec:3}.  Both {\MgI} and {\CaI} are
located outside of the forest, and are detected at $>5{\sigma}$
significance.  The {\SiI}~$\lambda$2515 detection is within the
forest, and cannot be confirmed by other {\SiI} transitions. However,
we expect that it is valid because of the precise alignment with the
$v$=-69~{\kms} cloud.  We thus conclude that there is more than
sufficient evidence that absorption in neutral species is detected,
making Cloud 1 in this system unique among other weak {\MgII} systems
\citep{Anand08}.

\section{Deriving Physical Conditions}
\label{sec:3}

\subsection{Oscillator Strengths and Saturation}
 The ratio of the {\MgII}~$\lambda$ 2803 to
{\MgII}~$\lambda$ 2796 equivalent width in Cloud 1 of the VLT
spectrum (Fig.~\ref{fig:clouds_labeled}) is not $0.5-0.7$ as
expected \citep{Anand08}, and as seen in Clouds 2-4, but is $0.84$; the weaker
member is considerably stronger than would be expected.  This ratio
implies either unresolved saturation of the line profile, or partial
covering of the quasar broad emission line region (BELR).  If the
profile were unresolved and saturated, profile fits to {\MgIIdblt} would
severely underestimate the column density and overestimate the Doppler
parameters of fits to the lines.  We consider this possibility in
defining a range of models to be considered.  

\subsection{Voigt Profile Fitting}

We initially perform Voigt profile fitting on the {\MgIIdblt} using
MINFIT \citep{Churchill03}.  We choose to optimize on {\MgII} (to
require Cloudy models to reproduce exactly the observed value) because
it is the only ion detected in all four clouds, and is the strongest
ion detected for this $z=0.4524$ absorber outside of the
{\Lya} forest.  The Doppler parameters
($b$) and column densities ($N$) given from the fit for Clouds 2-4 are
in Table~\ref{tab:Nb}, along with the redshifts of each
cloud.  The profile for Cloud 2 is asymmetric, and we find that a two
component fit provides a significantly better fit than does one
component.  These components are denoted as 2a and 2b in
Table~\ref{tab:Nb}. We note that the resolution of the spectrum,
$\sim 6.6$~{\kms}, is greater than the value of $b$ for four out of
the five Voigt profile components, implying that the clouds may not be
fully resolved.

The best fit to the Cloud 1 VLT spectrum is provided by a model with
column density $\log N({\MgII}) = 12.1~[{\rm cm}^{-2}]$ and Doppler
parameter $b=3.1 $~{\kms}, however this fit overproduces
{\MgII}~$\lambda$2796 and underproduces {\MgII}~$\lambda$2803, as
expected by the difference in $\frac{W_r({\MgII}\lambda
  2796)}{W_r({\MgII}\lambda 2803)}$ and the ratio of the oscillator
strengths of the two transitions.  
Fits to the other
clouds were adequate.  Because of the likelihood of unresolved
saturation, we increase $N$ until the ratio of {\MgIIdblt} equivalent
widths of the synthetic profile matches that of the observed
profile.  We also consider smaller Doppler parameters in our modeling process,
and for these values we adjust $N$ accordingly.

\subsection{Covering Factor}

We also consider the possibility of partial covering of the BELR of
HE0001-2340 to explain the observed {\MgIIdblt} equivalent widths.
The size of the broad emission line region of HE0001-2340 cannot be
calculated directly, as the spectrum is not flux calibrated.  However,
via comparison to a quasar of similar redshift and magnitude with a
flux calibrated spectrum, we may approximate the BELR size.  With
$z=2.28$ and $V=16.7$, HE0001-2340 may be compared to S5 0836+71 with
$z=2.172$ and $V=16.5$.
\citet{Benz07} gives the following relation for BELR size: 
$$
\log R_{BELR} [light days]=-22.198+0.539 \log \lambda L_{\lambda} (5100 \rm{\AA})
$$
 
With $\log \lambda L_{\lambda} \sim 47$, we estimate that the BELR
size of HE0001-2340 is $\sim 1.0$~pc.  Therefore, partial coverage of
the BELR requires a very small cloud size.  We explore this
possibility as an alternative to unresolved saturation, and perform VP
fitting to the {\MgII} doublet for Cloud 1, using a modified version
of MINFIT \citep{Churchill03} in which the covering factor,
$C_f({\MgII})$, is allowed to vary along with $N$ and $b$.  A covering
factor of $C_f({\MgII})=0.60\pm 0.10$ best fits the profile, as
determined by the $\chi^2$ minimization technique described in
\citet{Ganguly99}, and applied to a large doublet sample in
\citet{Misawa07}.  The $N({\MgII})$ and $b({\rm Mg})$ values for this
$C_f$ are $10^{12.1}$~{\cmsq} and 3.1~{\kms}, respectively.

\begin{figure}
\centering
\epsscale{1.0}
\includegraphics[scale=0.5]{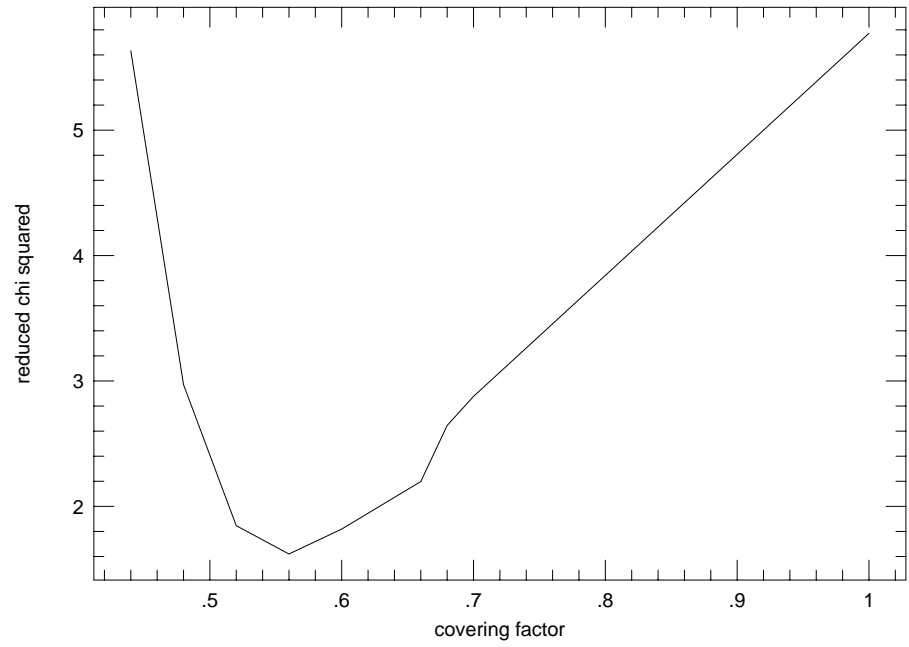}
\caption
{Reduced $\chi ^2$ value for Voigt profile fits of various covering factors to Cloud 1.  Covering factor values between $0.5$ and $0.7$ provide a significantly better fit than full coverage models.
\label{fig:cf}}
\end{figure}

Since we would not expect to find evidence for partial covering in
an absorber that is not intrinsic to a quasar, we examine how
significant an improvement $C_f({\MgII}) \sim 0.6$ provides as compared
to other possible values of the covering factor.  We force $C_f$
to have various other values and in each case find the best
$N$ and $b$ and compute the $\chi^2$ comparing the best model
to the data.  Figure~\ref{fig:cf} shows that a clear minimum
in $\chi^2$ occurs at $C_f({\MgII}) \sim 0.6$.

The $C_f$ measured from VP fitting is an "effective covering factor",
representing the absorption of the total flux at that wavelength, which
is a combination of flux from the quasar continuum source and BELR \citep{Ganguly99}.  The different
transitions for the intervening $z=0.4524$ system fall at different
positions relative to the quasar emission lines, and will
therefore absorb different relative fractions of continuum
and BELR flux.  In general, 
$C_f = \frac{C_c+{\rm W}C_{elr}}{1+{\rm W}}$, where
$C_c$ is the covering factor of the continuum, $C_{elr}$ is the
covering factor of the BELR,  and $C_f$ is the total covering factor
\citep{Ganguly99}.
The value of W, the ratio of the broad 
emission-line flux to the continuum flux at the wavelength of the narrow 
intervening absorption line ($F_{elr}/F_c$), 
can be determined for each transition using a low resolution spectrum
\citep{Tytler04}.  However, in order to calculate
the effective covering factor, $C_f$, we must make an assumption about
the relative covering factors of the continuum and BELR.

For {\MgII}, the value $C_f = 0.6$ can correspond to a range of
possible $C_c$, $C_{elr}$ pairs.  However, we know that
$C_c$ cannot be very small, as we see detections of
many transitions that are not superimposed on an emission line.
Since the continuum source is likely to be considerably smaller than
the BELR, it is most straightforward to assume that it is fully
covered and that the BELR is partially covered.  For the W value
measured for {\MgII} (${\rm W}=1.0$), and $C_c=1$, we then find
$C_{elr}=0.2$.
$C_{elr}$ values should be the same for all transitions, if their absorption
arises from 
the same cloud.
With this assumption, using the W values of each transition,
we can compute the effective covering factors as  
$C_f = \frac{1+0.2{\rm W}}{1+{\rm W}}$.  Table~\ref{tab:ew2} lists the $C_{f}$ values for all transitions
detected from Cloud 1.  Many of the detected neutral transitions
have $C_{f}$ values that are close to $1$, rendering their absorption
stronger relative to the {\MgII}, which only partially
absorbs a significant fraction of the incident flux due to its
position on a broad emission line.

\subsection{Cloudy Photoionization Modeling}
\label{sec:cloudy}

For each model, we begin with a column density, Doppler parameter,
and covering factor for {\MgII}.  For Cloud 1, we consider
a range of $N({\MgII})$, $b({\rm Mg})$ pairs
that are consistent with the data, including fits affected by unresolved
saturation.
Starting with the {\MgII} column density as a constraint, we used
the code Cloudy (version 07.02.01) to conduct photoionization models
\citep{Ferland98}.  For each cloud, we assume a plane-parallel slab
with the given $N({\MgII})$ and illuminate it with a Haardt-Madau
extragalactic background spectrum, including quasars and star-forming
galaxies (with an escape fraction of 0.1) at $z=0.4524$ \citep{HM96,
HM01}.  Given that the absorption is so weak in {\MgII}, it seems
unlikely that this absorber is housed in the midst of a galaxy
where the local stellar radiation field would be significant.
We initially run a grid of models for ionization parameters
$\log U=\log n_e/n_\gamma=-8.5$ to $-2.5$, and metallicities
$\log Z/Z_{\odot}= -3.0$ to $1.0$.  All models assume a solar abundance
pattern, unless otherwise stated.  The Cloudy output includes model column densities for all
transitions, as well as an equilibrium temperature, $T$, for the cloud.  The
turbulent component of the Doppler parameter
($b^2_{turb}=b({\MgII})^2-b_{therm}(\mathrm{Mg})^2$; where the thermal component
$b^2_{therm}=\frac{2kT}{m}$) is calculated from the equilibrium temperature and
observed $b({\MgII})$.  Given $b_{turb}$ and $b_{therm}$, the Doppler
parameter can be computed for all other elements.
After running Cloudy for all components from the Voigt profile fit,
the $b$ parameters and $C_f$ values are combined with the model
column densities to create a synthetic spectrum.
This model spectrum is compared to the data in order to constrain
the $\log U$ and $\log Z/Z_{\odot}$ values. The modeling method is the same as
that used by \citet{Ding05}, \citet{Lynch07}, and \citet{Misawa07}.

\section{Physical Conditions of the Absorbing Clouds}
\label{sec:4}

\subsection{Cloud 2}
\label{sec:4.1}

Cloud 2 can be fit with $-2.9<\log U<-3.3$; lower values overproduce
{\FeII}, while higher values underproduce {\FeII}.  Although there are
no Lyman series lines to use for direct constraints on the
metallicity, we find that if log$Z/Z_{\odot}< -0.3$, the {\MgI} is
overproduced. The temperature of the cloud is found to be $\sim 9000$K
for $\log Z/Z_{\odot}= -0.3$, with a size of tens of parsecs,
and a density $\sim 4$x$10^{-3}$ {\cc}.  Model parameters are listed in Table
\ref{tab:models}, where Clouds $2.1$ and $2.2$ represent the two components of Cloud 2, while the best model is shown in
Figure~\ref{fig:system_plot_unresolved}.

\subsection{Clouds 3 and 4}
\label{sec:4.2}
 Clouds 3 and 4 are quite weak, and are detected in the VLT
spectrum only because it has such a high S/N.  For Clouds 3 and 4, a
wide range of metallicities and ionization parameters provide an
adequate fit to the transitions covered by the spectrum. Low
ionization species other than {\MgII} are not detected at this
velocity, but because of the weakness of the lines, are also not
predicted by models with any reasonable ionization parameter, $\log U
> -7.5$. Coverage of higher ionization states would be needed to
further constrain the ionization parameter of these clouds.
Constraints for these clouds are summarized in
Table~\ref{tab:models}, where parameters are given for two acceptable
values of $\log U$ and log$Z/{Z_{\odot}}$ for each Cloud.

\subsection{Cloud 1}
\label{sec:4.3}

\subsubsection{Unresolved Saturation Model}

We run a grid of models with varying {\MgII} column densities, metallicities, and ionization parameters, as described in
section \S~\ref{sec:3}, for $b=3.1$ {\kms}, the Doppler
parameter given by a Voigt profile fit, which assumes the line
is resolved, unsaturated, and fully covered.  We find that detectable
amounts of {\FeI} are not produced in any model, and that {\FeII} is
always underproduced.  For smaller $b$ parameters we would expect that
${\FeI}/{\MgII}$ and ${\FeII}/{\MgII}$ would be larger because {\FeI}
and {\FeII} would be on the linear parts of their curves of growth
where the corresponding larger $N$ would affect their equivalent
widths.  In contrast, the {\MgII} would be on the flat part of its
curve of growth so that the increased $N$ would have little effect on
its equivalent width.  We therefore considered $N({\MgII})$, $b({\rm
Mg})$ pairs with $b=0.1-0.5${\kms} (giving $\log N({\MgII}) \sim 14$~{\cmsq})
and $\log U < -7.5$.  

An example model, with log$Z/{Z_{\odot}} =-1.0$, $\log U=-8.2$, and
$b=0.2$~{\kms} is shown in Fig.~\ref{fig:system_plot_unresolved}. Due
to the seemingly cold nature of this cloud, we opt to add dust grains
to the Cloudy \citep{Ferland98} simulations.  The primary flaw of this
model is the over-production of {\MgI}~$\lambda$2853 by $\sim 1$~dex
in column density; the apparent underproduction of some {\FeI} and
{\FeII} transitions may be attributed to {\Lya} forest contamination
of the observed profiles. Greater metallicities further overproduce
{\MgI}, and a small Fe abundance enhancement of $\sim 0.2$ dex is
needed to account for observed {\FeI} and {\FeII} profiles.

Constraints for this model are given in Table \ref{tab:models}, under Cloud
 $1^{a}$.
We find a size of 0.01-0.6 pc, T$<100$ K, and $500<n_H<1100$~[{\cc}]. With
a neutral hydrogen column density of $10^{18.7-20.8}$ {\cmsq} (sub-DLA to DLA
range), we find $0.72<
\frac{2N(H_2)}{2N(H_2)+N(HI)} < 0.91$.
$\rm{log}\frac{N(MgII)}{N(MgI)}\sim 1.0-1.3$ and
$\rm{log}\frac{N(FeII)}{N(FeI)} \sim 1.6-2.2$.

\subsubsection{Partial Covering Model}
\begin{figure}
\centering
\epsscale{1.0}
\plotone{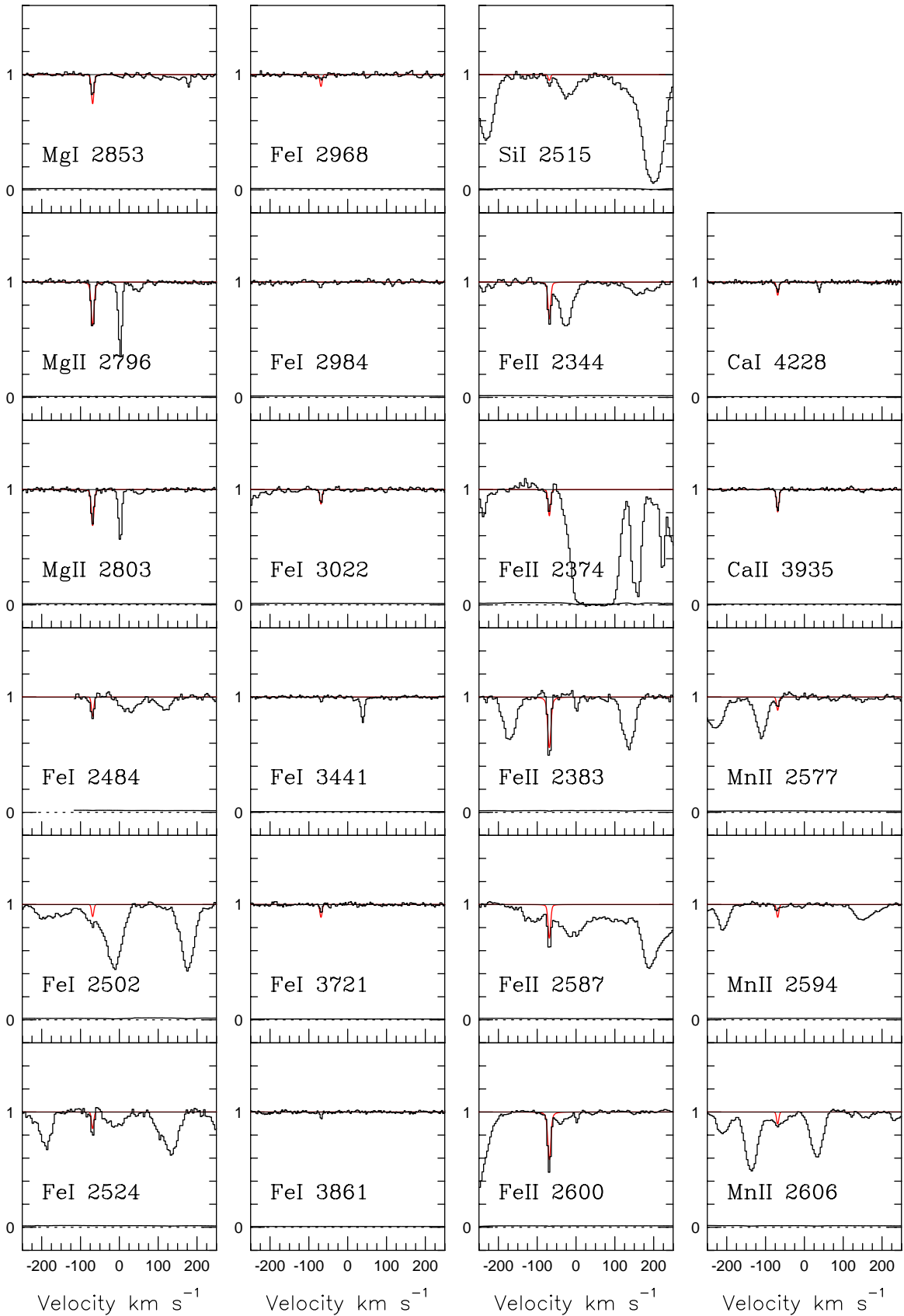}
\caption
{VLT spectrum (black) and photoionization model for partial covering case, with log$Z=-1.0$, log$U=-7.5$, log(N(MgII))=$14.65$, and $b=0.4${\kms} (red) 
\label{fig:system_plot_partial}}
\end{figure}

 We similarly find that a small $b$ ($<0.5~{\kms}$) is necessary to
reproduce the observed {\FeI} with partial coverage models.  An
example of such a model, with $b({\rm Mg}) = 0.4$~{\kms}, $\log U =
-7.5$, and $\log Z/Z_{\odot} = -1.0$ is given in
Fig.~\ref{fig:system_plot_partial}.  For this example model, the
observed {\FeI}, {\FeII}, {\CaI}, {\CaII}, {\MnII}, and {\SiI} are
adequately reproduced within the uncertainties.  The absorption at
the position of {\FeI}~$\lambda$2484 and $\lambda$2524 is not fully
produced, however the location of these transitions in the forest
makes contamination fairly likely.  The only discrepancy of this
model is the over-production of {\MgI}~$\lambda$2853 by $\sim 1$~dex
in column density.  In addition to this sample model, a range of
ionization parameters and metallicities provide a similar fit.  For
$\log Z/Z_{\odot} = -1.0$ and $\log U =-7.5$ the cloud size/thickness
is 0.2~pc, a size comparable to that of the quasar BELR.

The range of models that provide an adequate fit to this cloud have
extreme properties.  The ionization parameters for successful models
range from $-8.0$ to $-7.0$, implying densities of $30 < n_H <
1100$~{\cc}. For the range of possible metallicities, the equilibrium
temperatures are low, $< 50$~K.  The neutral hydrogen column densities,
$\log N({\HI})=18.8-19.9$~{\cmsq}), are in the sub-DLA range, while $0.76<
\frac{2N(H_2)}{2N(H_2)+N(HI)} < 0.94$ . Cloud
properties for acceptable models are summarized in in
Table~\ref{tab:models} under Cloud $1^b$, while a sample model is
plotted in Figure~\ref{fig:system_plot_partial}.

\section{Discussion}
\label{sec:5}

The $z=0.4524$ system toward HE0001-2340, shown in Figures
~\ref{fig:clouds_labeled} and~\ref{fig:system_plot_unresolved},
significantly differs from typical weak {\MgII} absorption line
systems, due to the detection of low ionization states in one of four
main absorption components.  We consider the environments of these
clouds in the context of known structures that could produce these
absorption signatures.
 
\subsection{Clouds 2, 3, and 4}

Cloud 2 has conditions similar to those of the 100 weak {\MgII}
systems modeled by \cite{Anand08}, with $N({\MgII})=10^{12.5}$~{\cmsq},
 $N({\FeII})=10^{11.85}$~{\cmsq}, a size of tens of parsecs, a density of 
$\sim 0.004 {\cc}$, and a temperature of
$\sim 9000$K (Table~\ref{tab:models}).  
Clouds 3 and 4 also fall within the
range of properties exhibited in the
\cite{Anand08} sample, and although their properties are not
well-constrained by models, they appear to be of a temperature and
density typical of the weakest {\MgII} absorbers studied to date.
Detection of such low $N({\MgII})$ absorbers is limited to very high
S/N spectra, suggesting that such clouds may often go undetected near
what are perceived as single-cloud weak {\MgII} absorbers.

\cite{Rigby} divides weak {\MgII} absorbers at $\log [N({\FeII})/N({\MgII})]=-0.3$
into iron-rich and iron-poor subcategories.  While it is not possible
to classify Clouds 3 and 4, Cloud 2, with log(N(FeII)/N(MgII))=-0.62,
falls into the ``iron-poor'' subcategory, with corresponding $\log U
\sim -3.1$.  Clouds associated with superwind condensations could be 
responsible for this iron-poor environment, as type II supernovae
driving the wind will lead to ${\alpha}$-enhancements, as they build up
high metallicities. Such systems may be predecessors to the high
metallicity ($>0.1~Z_{\odot}$) {\CIV} absorbers of \cite{Schaye07},
which have sizes $\sim 100$~pc.  With radii less than the Jean's
length for self-gravitating clouds, these high metallicity {\CIV}
clouds are likely to be short-lived, with lifetimes on the
order of the time-scale for free expansion, $\sim 10^7$ years.  In comparison,
Cloud 2 would have a lifetime on the order of $\sim 10^6-10^7$ years in its present state, which could precede its {\CIV} cloud phase.

\subsection{Cloud 1}

Based solely upon its detected absorption features, Cloud 1 is an
anomaly.  {\FeI} has only been detected along a fraction of Galactic
sightlines passing through molecular clouds \citep{Welty03}.  These
sightlines pass through clouds that are very cold ($<100$K), and have
densities ranging from $\sim 10-300$~{\cc}, on par with the results
given by photoionization models of Cloud 1.

Partial covering and unresolved saturation models of Cloud 1 in the
VLT spectrum provide similar constraints to its properties as those
dense Galactic clouds.  Both
require a narrow {\MgII}2796{\AA} profile, with $b<0.5$~{\kms}, to
reproduce the observed {\FeI}, suggesting that regardless of whether
the system is partially covering the BELR, it is unresolved.  We note
that the other properties of Cloud 1 given by Cloudy are similar in
both models, as shown in Table~\ref{tab:models}.  Such models are not
mutually exclusive, as it would be impossible to distinguish partial
covering from unresolved saturation based on a comparison of the
members of the {\MgIIdblt} doublet. With either model, the size of the
absorber is likely to be on the order of the QSO BELR size at the
distance of the absorber (parsec-scale), thus partial covering is not
unlikely. It is important to note that the covering factors of
different transitions depends upon their position on broad emission
lines, a concept that is key to intrinsic absorption line studies,
as partial covering by clouds in the immediate vicinity of the quasar
is common.

\subsubsection{Detection Rates}
\label{sec:5.2.1}
In a sample of 81 VLT QSOs, with a redshift path length of $\sim 75$
\citep{Anand07}, we have found only one system with detected {\FeI}.  However,
considering the small size of this object, even this one detection is
significant, suggesting that there
may be a significant cross section covered by these objects, which lie
in the outskirts of galaxies and remain undetected.  Assuming the
same average redshift pathlength per quasar, $\Delta z = 0.93$, and a
Poisson distribution for the detections, we find that there is a
$50\%$ chance of one or more such detections in a sample of 56 more
quasars, and a $95\%$ chance of one or more detections in a sample of
243 more quasars (which corresponds to 332 more weak absorption line
systems).  The one {\FeI} system that was detected in a pathlength
of $75$ yields $dN/dz=0.013$, and there is a probability of 95\% that
$dN/dz > 0.0044$ (based on one detection in a pathlength of $243 \times 0.93$).

\cite{Glenn08} finds that, for a range of Schechter luminosity function
parameters, a statistical absorption radius of $43 < R_x < 88$~kpc
would explain the observed $dN/dz \sim 0.8$ of strong {\MgII} absorbers
($W_r(2796) > 0.3$~{\AA}) at $0.3 < z < 1.0$.  To consider the radius specifically
at the redshift of the $z=0.4524$ absorber,  we scale those results
considering that \cite{Nestor05} find $dN/dz \sim 0.6$ for strong {\MgII} 
absorbers at $z \sim 0.45$.  This implies a corresponding absorber halo
radius of 24-49~kpc, assuming a covering factor of order unity.   From the
estimated $dN/dz$ of {\FeI} absorbers, we estimate that approximately $2\%$
of the area covered by strong {\MgII} absorbers ($1810$-$7540$ kpc$^2$)
is covered by such small molecular clouds.
These clouds are likely to take the form of filaments or thin sheets around
many galaxies.

It is therefore of interest to consider what the covering factor would
be of molecules in the Milky Way halo, if we were to observe it from
outside the halo.  The covering factor of 21-cm high velocity clouds,
to a limit of $N(HI) = 7 \times 10^{17}$~{\cmsq}, was measured to be
37\%, looking out from our vantage point.  Although molecules have
been detected in several HVCs \citep{Richter99,Sembach01,Richter01a,
Richter01b}, the fraction of HVC sightlines
with detected H$_2$ is small, e.g. 1/19 in the FUSE survey of
\citet{Wakker06}.  Nonetheless, looking through the Milky Way from
outside, we might expect roughly $2 \times 37$\%$\times (1/19) \sim 4$\%
as the covering factor for molecular absorption, consistent with the
2\% halo covering factor that we estimated for {\FeI} absorbers at $z \sim 0.45$.

An alternative to a large fraction of galaxies producing {\FeI} absorbers
with small individual covering factors is a small population of galaxies that
which have larger individual covering factors.  One possibility is starburst
galaxies with superwinds, since the neutral species {\NaI} is
commonly detected in absorption
through these objects \citep{Heckman00}.  Nearby starbursts are found to
have strong {\NaI} absorption over regions $1$-$10$~kpcs in extent.  Weaker
{\NaI} absorption, consistent with its expected strength for Cloud 1
in the $z=0.4524$ absorber, would be
detected over larger areas.  Also, starbursts are more
common at $z\sim0.45$ than at the present.  These factors could combine to
lead to a significant covering factor of {\NaI} (and therefore {\FeI}) absorption
from starburst winds at the redshift of the $z=0.4524$ absorber.

\subsubsection{Origin}
\label{sec:5.2.2}
The model gives neutral hydrogen column densities of the absorber that
suggest a sub-DLA or DLA environment, with $10^{18.8}< N({\HI})<
10^{20.8}$~{\cmsq}.  Although H$_2$ is frequently detected in higher
redshift systems, the molecular hydrogen fraction is only $<10^{-6}$ in
$58-75\%$ of the high-z DLAs surveyed by \cite{Ledoux03}, and does not
rise above $\sim 0.03$ for sub-DLAs and DLAs.  \cite{Hirashita03}
perform simulations of high-z DLAs, and find that the area of a region
that would produce DLA absorption that contains molecules is a very
small fraction of its overall area.  They note that in high
ultraviolet background, the molecular fraction can be $>10^{-3}$,
which is large enough to fuel star formation.  Small, dense ($n_H \sim
10-1000$), cold ($<100$K) absorbers, with significant molecular
fractions, like the $z=0.4524$ absorber toward HE0001-2340, may occur
over a small fraction of structures that give rise to sub-DLAs and
DLAs, but are rarely intersected by QSOs due to their relatively smaller area.

While extremely high molecular hydrogen fractions seem to be rare in
sub-DLAs and DLAs, up to $53\%$ of total hydrogen mass was found to be
molecular in a survey of
\cite{Leroy09}, which examined CO emission from 18 nearby galaxies.  
In these detections, the H$_2$ mass can be as high as 0.1 times the stellar
mass.  The ratio of CO $J=2 \rightarrow 1$ intensity to CO $J=1 \rightarrow 0$
intensity suggests that the emitting gas is optically thick with an
excitation temperature of $\sim 10$K.  Although the exponential scale
lengths of the survey targets range only from 0.8-3.2 kpc, they note
that with increased sensitivity, objects at larger distances could be
detected.  

Similar high resolution CO observations of structures likely belonging
to the Milky Way by \cite{Heithausen02, Heithausen04} show fractalized
clumpuscules at high galactic latitudes, with sizes of $\sim 100$ AU,
masses $\sim 10^{-3} M_{\odot}$, $n($H$_2)=1000-20000~{\cc}$, and {\HI}
column densities of $\sim 10^{20}~{\cmsq}$; he notes that such small structure
would not be visible in the inner galaxy, due to overcrowding in
normal molecular gas.  A previous study of high galactic latitude
molecular clouds via stellar absorption lines by
\cite{Penprase93} likewise found high molecular hydrogen fractions, 
high densities, and small sizes.

\citet{Pfenniger94a} and \citet{Pfenniger94b}
suggest such small structures may make up a significant percentage of
halo matter, and propose that such structures may account for a
fraction of missing baryonic dark matter.  These structures are believed
to be formed by Jeans unstable and isothermal clouds, which fragment
into smaller cloudlets.   An alternative to this
fractalized ISM model is explored through simulations by \cite{Glover07}.
They study H$_2$ formation in turbulent gas, predicting that H$_2$
should rapidly form in any dense turbulent atomic cloud (within 0.05
Myr), regardless of whether or not the cloud is gravitationally bound
or whether it is magnetic subcritical or supercritical; up to $40\%$
of the initial atomic hydrogen may be converted to H$_2$ in this process.  They
find that regions of low density ($n<300 {\cc}$) contain more H$_2$
than expected in gas in photodissociation equilibrium due to turbulent
transport of H$_2$ from higher density ($n>1000 {\cc}$) areas.

Hydrodynamic simulations by \cite{Fujita09} of the blowout of
starburst-driven superbubbles from the molecular disk of ULIRGs also
show similar small-scale clumps.  Thin dense shells are created as the
bubbles sweep up dense disk gas; Rayleigh-Taylor instabilities then
cause these shells to fragment.  Clumpiness is seen down to the limits
of the simulation resolution of 0.1 pc.  These clumps contain high
{\NaI} mass, and have $N({\rm H}) \sim 10^{21}$~{\cmsq}.

As small molecular structures have been observed both in the galactic
neighborhood (on the order of hundreds of AU),and in
hydrodynamic simulations (to the limit of their resolutions) one would
expect there to exist such structures beyond the local universe, at higher
redshifts.  Because these structures are far too small to detect via
imaging of extragalactic environments, absorption lines serve as the
only probe of small-scale structure of distant galaxies.  The $z=0.4524$
absorber may be a first glimpse of such a cold, dense cloud at an
earlier epoch.

\subsubsection{Iron content}
Cloud 1, with $\log$({\FeII}/{\MgII})$\sim 0.5$, is an extreme case of
an iron-rich absorber.  Iron-rich systems discussed in \cite{Rigby} were generally found to
have high metallicities ($>0.1 Z_{\odot}$), $\log U<-4.0$ ($n_H >0.09$~{\cc}),
 and small sizes ($N$({\HI})+$N$({\HII})$<10^{18}$~{\cmsq}, $R<10$~pc).
Solar abundances, and not $\alpha$-enhancement, are needed to explain
their similar {\FeII} and {\MgII}
column densities, which implies Type Ia event-enrichment of the gas
\citep{Timmes95}.  Type Ia enrichment requires a $\sim 1$ billion year
delay from the onset of star formation before the elements produced 
enter the interstellar medium, which could be consistent with gas 
trapped in potential wells of dwarf galaxies, or with intergalactic 
star-forming structures in the cosmic web \citep{Rigby, Nikola06}.  Rapid 
condensation of turbulent gas from a supernova could explain the small
size, high molecular content, and high iron abundance observed in Cloud 1.
 
\subsection{The System}
Weak {\MgII} absorption systems are not often found within 30 kpc of
luminous star-forming galaxies, with $L>0.05 L_*$ \citep{CLB98,
Churchill05, Mil06}; the astrophysical origins of such systems have
not been identified, although they may include extragalactic high
velocity clouds \citep{Anand07}, dwarf galaxies \citep{Lynch06},
material expelled by superwinds in dwarf galaxies \citep{Zonak04,
Stocke04, Keeney06}, and/or massive starburst galaxies and
metal-enriched gas in intergalactic star clusters \citep{Rigby}.
\cite{Schaye07} suggest that at intermediate redshifts ($z \sim 2-3$),
weak {\MgII} clouds may have been ejected from starburst
supernova-driven winds during an intermediate phase of free expansion,
prior to achieving equilibrium with the IGM.  Supernova-driven winds
are believed to have a multi-phase structure, with a cold component
($T \sim 100$K) detected in {\NaI} \citep{Heckman00, Rupke02, Fujita09}.
A warm neutral phase ($T \sim 10000$K) may surround this component
\citep{Schwartz06}.  
This gas will likely fragment through
hydrodynamical instabilities as it moves through the halo of the
galaxy, and would appear initially as weak {\MgII} absorption, and
later as weak {\CIV} absorption associated with {\HI} lines in the
{\Lya} forest \citep{Zonak04,Schaye07}.

To account for the differing {\FeII}/{\MgII} observed in the different clouds of
the same system, it seems necessary to propose different stellar
populations in the vicinity of the absorbing structure.  
Cloud 1, due
to its iron-rich nature, must be enhanced by a Type Ia supernova,
implying the gas does not originate from a very young stellar population.
The density and temperature of this gas, comparable to that of Milky
Way molecular clouds, suggests that it is a potential site of
star formation.  Cloud 2, in contrast, is likely a remnant of a
supernova from a massive star recently formed in its vicinity, or
in a superwind driven by a collection of massive stars, as
suggested by its ${\alpha}$-enhancement.  
It is not surprising that we would see a mix of different processes
and stellar populations in an absorption line system.
Some combination of dwarf galaxies, high velocity clouds, superwind
and supernova shell fragments, and tidal debris could give rise
to such variety.  In such a scenario, clouds like Cloud 2 would
commonly be grouped together into systems, sometimes with Fe-rich
clouds, but something as extreme as Cloud 1 would be rare.

\section{Conclusions}
\label{sec:6}
Although {\FeI} detections are extremely rare in extragalactic
absorbers, photoionization models of Cloud 1 suggest that its small
size ($<1$pc) may cause many such absorbers to go undetected; we
estimate that two percent of the areas of $z\sim0.45$ halos 
(24-49~kpc in radius) should be
covered by such objects.
With cold temperatures ($<100$K), high densities
($30<n_H<1100$), and a large molecular hydrogen fraction
($72-94\%$), the properties of Cloud 1 are similar to the dense, small
Milky Way clouds at high galactic latitudes observed in CO by
\cite{Heithausen02, Heithausen04} and \citet{Penprase93}, suggesting that pockets
of gas like Cloud 1 may be analogs of these Milky Way clouds in the
halos of other galaxies.

Small scale clumps, with sizes of hundreds of AU, are predicted to
exist both as part of the fractalized ISM in the halos of galaxies
\citep{Pfenniger94a, Pfenniger94b} and as condensates in turbulent gas
\citep{Glover07, Fujita09}.  A Type-Ia supernova may have been responsible for
both the turbulence and observed iron abundance in Cloud 1.

In contrast, the ${\alpha}$-enhancement observed in Cloud 2, with its
low-Fe content, is suggestive of an origin in a Type II
supernova-driven superwind. Multiple-cloud weak {\MgII} absorption line systems
are generally thought to originate in lines of sight that pass through
more than one dwarf galaxies, through sparse regions of
luminous galaxies at high impact parameters, or in gas-poor
galaxies.  In this case, one of the four components happens to
be a relatively rare type of cold and dense weak {\MgII} absorber.

Via imaging of the HE0001-2340 field it may be possible to
further constrain the absorption origin, perhaps by identifying a
host galaxy or galaxies, or by finding a nearby galaxy whose environment
is being sampled by these absorbers.  It is also of interest to
find another similar {\FeI} cloud, by searching large numbers of
high-$S/N$, high-resolution quasar spectra.  Perhaps in another
case, the sightline would be clear enough to allow access to the
Lyman series lines, so that its metallicity could be directly measured,
or to higher ionization absorption lines so that any surrounding, lower
density gas phases could be identified.

\acknowledgements{This research was funded by the National Science
Foundation (NSF) under grant AST-04-07138, by the Graduate Research
Fellowship Program, and through the REU program.  We also received
support from NASA under grant NAG5-6399 NNG04GE73G. Support by NSF 
(0607028 and 0908877) and NASA (07-ATFP07-0124) are gratefully 
acknowledged by GJF. We express our
gratitude to the ESO for making the VLT/UVES spectra available for
the public and for providing the ESO-MIDAS UVES reduction pipeline
under a GNU General Public License.}

\clearpage

\clearpage

\clearpage

\clearpage

\clearpage
\begin{deluxetable*}{ccccccc}
\tabletypesize{\scriptsize}
\setlength{\tabcolsep}{0.03in}
\tablecaption{Equivalent widths and equivalent width limits of selected transitions \label{tab:ew}}
\tablewidth{0pt}
\tablehead{
\colhead{Cloud}            &
\colhead{$z$}              &
\colhead{$W_r({\MgII} 2796) $({\AA})}  &
\colhead{$W_r({\MgII} 2803) $({\AA})}  &
\colhead{$W_r({\FeII} 2383)$({\AA})}		   &
\colhead{$W_r({\FeII} 2600)$({\AA})}   &
\colhead{$W_r({\MgI} 2853)$({\AA})}
}
\startdata
\multicolumn{7}{c}{}\\ 
$1$ & 0.452061 & $0.0377 \pm 0.0006$ & $0.028 \pm 0.001$ & 0.028 $\pm$ 0.007 &  0.0392 $\pm$ 0.0006 & 0.0128 $\pm$ 0.0006 \\ \hline
$2$ & 0.452387 & 0.0761 $\pm$ 0.0004 & $0.04551 \pm 0.0009$ &  $0.010 \pm  0.001$ & $0.0079 \pm 0.0008$  &  $<0.005$ \\ \hline
$3$ & 0.452622 & 0.019 $\pm$ 0.001 & $0.008 \pm 0.001$ & $<0.004$ & $<0.004$ & $<0.003$ \\ \hline
$4$ & 0.454864 & 0.009 $\pm$ 0.001 & $0.002 \pm 0.001$ & $<0.003$ & $<0.04$ & $<0.003$

\enddata 

\end{deluxetable*}

\clearpage

\begin{table}[h!b!p!]
\caption{Cloud redshifts and VP fit-derived column density and Doppler parameters, assuming full coverage
\label{tab:Nb}} 
\begin{tabular}{|l|l|l|l|l|} \hline
Cloud & z & log(N({\MgII}2796)) & b ({\kms}) \\ \hline
1 & 0.452061 & $12.10 \pm 0.04$ & $3.1 \pm 0.1$\\ \hline
$2^a$ & 0.452387 & 11.8 $\pm$ 0.4 & 5 $\pm$ 2 \\ \hline
$2^b$ & 0.452410 & 12.4 $\pm$ 0.1 & 2.77 $\pm$ 0.3 \\ \hline
3 & 0.452622 & 11.67 $\pm$ 0.02 & 15 $\pm$ 1 \\ \hline
4 & 0.454864 & 11.30 $\pm$ 0.05 & 7 $\pm$ 1\\ \hline
\end{tabular}
\end{table}

\clearpage
\begin{table}[h!b!p!]
\caption{Oscillator strengths and equivalent widths for transitions detected in Cloud 1, the ratio of broad emission-line flux to continuum flux, W, and the covering factors, $C_f$ for partial covering model
\label{tab:ew2}}
\begin{tabular}{|l l|l|l|l|l|} \hline
Ion & Transition & $f_{lu}$ & $W_r$ ({\AA}) & W & $C_f$ \\ \hline
Mg II & 2796 & 0.6123 & 0.0377 $\pm$ 0.0006 & 1.0 & 0.60 \\ \hline
Mg II & 2803 & 0.3054 & 0.028 $\pm$ 0.001 & 1.0 & 0.60 \\ \hline
Mg I & 2853 & 1.810000 & 0.0186 $\pm$ 0.0008 & 0.33 & 0.80 \\ \hline
Fe I & 2484 & 0.557000 & 0.013 $\pm$ 0.001 & 0.21 & 0.86 \\ \hline
 & 2502 & 0.049600 & $<0.097$ & 0.11 & 0.92 \\ \hline
 & 2524 & 0.279000 & 0.018 $\pm$ 0.001 & 0.11 & 0.92 \\ \hline
 & 2968 & 0.043800 & 0.011 $\pm$ 0.001  & 0 & 1.00 \\ \hline
 & 2984 & 0.029049 & 0.0045 $\pm$ 0.0008 & 0 & 1.00 \\ \hline
 & 3022 & 0.10390 & 0.011 $\pm$ 0.001 & 0 & 1.00 \\ \hline
 & 3441 & 0.02362 & 0.0033 $\pm$ 0.0006 & 0.86 &  0.63 \\ \hline
 & 3721 & 0.04105 &  0.0061 $\pm$ 0.0007 & 0.27 & 0.83 \\ \hline
 & 3861 & 0.02164 & 0.0048 $\pm$ 0.0004 & 0.07 & 0.95 \\ \hline
Si I & 2515 &  0.162000 & 0.017 $\pm$ 0.001 & 0.11 & 0.92 \\ \hline
Fe II & 2344 & 0.109700 & 0.036 $\pm$ 0.001 & 0.57 & 0.71 \\ \hline
 & 2374 & 0.02818 & 0.014 $\pm$ 0.002  & 0.43 & 0.76 \\ \hline
 & 2383 & 0.3006 & 0.040 $\pm$ 0.001 & 0.43 & 0.76 \\ \hline
 & 2587 & 0.064570 & 0.078 $\pm$ 0.001 & 0.21 & 0.86 \\ \hline
 & 2600 & 0.22390 & 0.0569 $\pm$ 0.0008 & 0.21 & 0.86 \\ \hline
Ca I & 4228 & 1.7534 & 0.0051 $\pm$ 0.0008 & 0.18 & 0.88 \\ \hline
Ca II & 3935 & 0.6346 & 0.0179 $\pm$ 0.0004 & 0.13 & 0.91 \\ \hline
Mn II & 2577 & 0.03508 & 0.011 $\pm$ 0.001 & 0.11 & 0.92 \\ \hline
 & 2594 & 0.2710 & 0.001 $\pm$ 0.0002 & 0.21 & 0.86 \\ \hline
 & 2606 & 0.1927 & $< 0.028$ & 0.21 & 0.86 \\ \hline
\end{tabular}
\end{table}
\clearpage
\begin{landscape}
  \begin{deluxetable}{ccccccccccccccccc}
\tabletypesize{\tiny}
\setlength{\tabcolsep}{0.0in}
\tablecaption{Sample models that fit the system.  
 \label{tab:models}}
\tablewidth{0pt}
\tablehead{
\vspace{2mm} \\
\colhead{Cloud}            &
\colhead{$\log (Z/Z_{\odot})$}  &
\colhead{$\log (U)$}         &
\colhead{$n_{H}$}          &
\colhead{Size}             &
\colhead{$T$}              &
\colhead{$N$(HI)}   &
\colhead{$N$(HII)}    &
\colhead{$N$(H$_2)$}	   &
\colhead{$N_{tot}$(H)} &
\colhead{$\frac{2N(H_2)}{2N(H_2)+N(HI)}$} & 
\colhead{$N$(MgI)} &
\colhead{$N$(MgII)} &
\colhead{$N$(FeI)} &
\colhead{$N$(FeII)} &
\colhead{$b$(Mg)}          & \\
\colhead{}                 &
\colhead{}                 &
\colhead{}		   &
\colhead{(cm$^{-3}$)}      &
\colhead{(pc)}            &
\colhead{(K)}              &
\colhead{(cm$^{-2}$)}      &
\colhead{(cm$^{-2}$)}      &
\colhead{(cm$^{-2}$)}      &
\colhead{(cm$^{-2}$)}      &
\colhead{}		&
\colhead{(cm$^{-2}$)}      &
\colhead{(cm$^{-2}$)}      &
\colhead{(cm$^{-2}$)}      &
\colhead{(cm$^{-2}$)}      &
\colhead{(\kms)}           &\\
\colhead{(1)}              &
\colhead{(2)}              &
\colhead{(3)}              &
\colhead{(4)}              &
\colhead{(5)}              &
\colhead{(6)}              &
\colhead{(7)}              &
\colhead{(8)}              &
\colhead{(9)}              &
\colhead{(10)}             &
\colhead{(11)}             &
\colhead{(12)}             &
\colhead{(13)}             & 
\colhead{(14)}              &
\colhead{(15)}              &
\colhead{(16)}              & \\
}
\startdata
\multicolumn{16}{c}{Model} \\
\hline

$1^{a}$  & $<-1.0$ & $-8.5~\rm{to} -7.5$ & 500-1100 & 0.01-0.6 & $<100$ & $10^{18.5-20.8}$ & $10^{15.9-16.5}$ & $10^{19.2-21.1}$ & $10^{19.3-21.3}$ & 0.72-0.91 & $10^{12.8-13.1}$ & $10^{14.1}$ & $10^{12.4-13.0}$ & $10^{14.6} $& $<0.5$ \\ \hline
$1^{b}$ & $<-0.3$ & $-8.0~\rm{to} -7.0$ & 30-1100 & 0.08-0.19 & $<50$ &  $10^{18.8-19.9}$ & $10^{16.1-16.3}$ & $10^{19.0-20.8}$ & $10^{19.2-20.9}$ & 0.76-0.94 & $10^{13.2-13.4}$ & $10^{14.7}$ & $10^{13.0-13.4}$ & $10^{15.3}$ & $<0.5$ \\ \hline 
$2.1$ & $-0.3$ & $-3.1$ & 0.004 & 15 & 9200 & $10^{15.14}$ & $10^{17.30}$ & $10^{5.10}$ & $10^{17.30}$ & 0 & $10^{9.83}$ & $10^{11.79}$ & $10^{7.25}$ & $10^{11.17}$ & 0.4\\ \hline
$2.2$ & -0.3 & $-3.1$ & 0.004 & 57 & 9100 & $10^{15.71}$ & $10^{17.87}$ & $10^{5.68}$ & $10^{17.88}$ & 0 & $10^{10.40}$ & $10^{12.37}$ & $10^{7.83}$ & $10^{11.75}$ & 2.77 \\ \hline
$3^{a}$  & $-1.0^*$ & $-7.5$ & 110 & 0.0004 & $ 3000$ & $10^{17.13}$ & $10^{15.84}$ & $10^{12.60}$ & $10^{17.15}$ & 0 & $10^{10.62}$ & $10^{11.65}$ & $10^{7.71}$ & $10^{11.60}$ & $14.53$ \\ \hline
$3^{b}$ & $-0.3^*$ & $-3.1$ & 0.004 & 110 & 9200  & $10^{15.01}$ & $10^{17.17}$ & $10^{4.98}$ & $10^{17.18}$ & 0 & $10^{9.70}$ & $10^{11.67}$ & $10^{7.12}$ & $10^{11.05}$ & 14.53 \\ \hline
$4^{a}$ & $-1.0^*$ & $-7.5$ & 110 & 0.0002 & 3250 & $10^{16.78}$ & $10^{15.52}$ & $10^{12.07}$ & $10^{16.80}$ & 0 & $10^{10.17}$ & $10^{11.30}$ & $10^{7.35}$ & $10^{11.25}$ & 6.00 \\ \hline
$4^{b}$ & $-0.3^*$ & $-3.1$ & 0.004 & 5 & 9200 & $10^{14.64}$ & $10^{16.80}$ & $10^{4.60}$ & $10^{16.80}$ & 0 & $10^{9.33}$ & $10^{11.29}$ & $10^{6.75}$ & $10^{10.67}$ & 6.00 \\

\enddata
\vspace{2mm} \\
\tablenotetext{*}{a wide range of metallicities and ionization parameters provide reasonable fits to these clouds}
\end{deluxetable}
\clearpage
\end{landscape}
\end{document}